\documentclass[a4paper,fleqn]{cas-sc}

\usepackage[authoryear,longnamesfirst]{natbib}
\newtheorem{example}{Example}

\newtheorem{remark}{Remark}

\newtheorem{problem}{Problem}

\usepackage{subfigure} 

\def\tsc#1{\csdef{#1}{\textsc{\lowercase{#1}}\xspace}}
\tsc{WGM}
\tsc{QE}

\begin{document}
\let\WriteBookmarks\relax
\def\floatpagepagefraction{1}
\def\textpagefraction{.001}

\shorttitle{}    

\shortauthors{}  

\title [mode = title]{Multi-Agent Control Synthesis from Global Temporal Logic Tasks with Synchronous Satisfaction Requirements}  



%

\author[1]{Tiange Yang}
\ead{tgyang@sjtu.edu.cn}
\credit{Investigation, Software, Validation, Writing - original draft,  Writing - review \& editing}
\affiliation[1]{organization={Department of Automation, Shanghai Jiao Tong University},
            city={Shanghai},
            postcode={200240}, 
            country={China}}

\author[1]{Yuanyuan Zou}
\cormark[1]
\ead{yuanyzou@sjtu.edu.cn}
\credit{Investigation, Methodology, Writing - review \& editing}
\affiliation[2]{organization={Department of Chemical \& Materials Engineering, University of Alberta},
            city={Edmonton},
            postcode={T6G 1H9}, 
            country={Canada}}

\author[2]{Jinfeng Liu}
\ead{jinfeng@ualberta.ca}
\credit{Conceptualization, Supervision}
\author[1]{Shaoyuan Li}
\ead{syli@sjtu.edu.cn}
\credit{Conceptualization, Supervision}
\author[1]{Xiaohu Zhao}
\ead{xiaohuz@sjtu.edu.cn}
\credit{Investigation, Software, Validation}

\cortext[1]{Corresponding author}

\begin{abstract}
This paper addresses the multi-agent control problem in the context of global temporal logic tasks, considering agents with heterogeneous capabilities. These global tasks involve not only absolute and relative temporal and spatial constraints, but also group behaviors, including task completion times, agent capabilities, and task interdependencies such as the need for synchronous execution. The global tasks are formally formulated into global signal temporal logic (STL) formulae, and a synchronous robustness metric is designed to evaluate the synchronization quality with real values. A mixed-integer linear programming (MILP) encoding method is further proposed to realize task-satisfied motion planning with high synchronicity and minimum control efforts. The encoding method uses a logarithmic number of binary variables to fully capture synchronous robustness, leading to only linear computational complexity. Simulations are conducted to demonstrate the efficiency of the proposed control strategy.
\end{abstract}





\maketitle

\section{Introduction}
Multi-agent systems can serve modern society in a variety of ways, ranging from pure entertainment to practical applications such as industrial systems and traffic networks \cite{qin2016recent,10207769,10458342,10574325}.
In recent years, increasing attention has been paid to control synthesis under complex temporal logic tasks, where strict time and logic constraints are imposed on agent behaviors \cite{9687668,9929325}. For example, \emph{always avoid collision with obstacles, do not cross into region A until region B is visited, and eventually visit regions A and C within given time window $[t_1,t_2]$}. Due to the expressivity and similarity to natural languages, temporal logic specifications \cite{bartocci2018specification, STL_intro}, such as  linear temporal logic (LTL) and signal temporal logic (STL), have been widely used to formalize tasks for control systems. Multi-agent systems under temporal logic tasks can be bottom-up or top-down \cite{lindemann2019coupled}. In the former, each agent is assigned some local and coupled tasks individually \cite{gundana2021event, yang2023distributed}, while in the latter, temporal logic tasks are not imposed on specific agents, but rather the agents collaborate to achieve a global task \cite{bai2022hierarchical}, which is considered in this work.

In many multi-agent systems, agents have diverse capabilities, and certain global tasks require robots of different types to collaborate. Consider a specification for a multi-robot surveillance mission as follows: \emph{Region A must be monitored every 10 minutes by one robot equipped with a high-resolution camera. In region B, two ground robots need to be present simultaneously within time window $[t_1,t_2]$. Additionally, region C requires a total of 10 visits by either a ground or an aerial robot throughout the mission}. The complexity of these temporal logic missions rises due to the temporal and logical constraints, and also the frequency and agent capabilities required. Very recent work has focused on the development of control synthesis algorithms specifically tailored for such situations. In \cite{sahin2019multirobot}, global tasks, such as determining the number of specific agents required for a given LTL task, were considered, and an optimization-based method was proposed to generate individual trajectories that collectively satisfy specifications. In \cite{luo2022temporal}, a global LTL formula $\mbox{LTL}^{\mathcal{X}}$ was explored to enable tasks that require agents of various types to visit different regions in sequence. However, these LTL-based specifications operate on untimed sequences and are unable to capture precise timing constraints. To address this limitation, STL-based global tasks were considered in \cite{leahy2021scalable}, which indicates the number of agents with specific capabilities required to reach a region and stay for a certain period. Task-based coordination with online replanning was proposed to handle agent attrition using mixed integer linear programming (MILP). Similar global tasks were considered in~\cite{buyukkocak2021planning}, with the extension to involve integral predicates. Tasks in \cite{liu2023robust} considered the number of agents needed to satisfy a specific STL task. Exponential robustness was designed accordingly to quantify how strongly the task is satisfied from a spatial perspective, which is sound and mask-eliminated, and a robust algorithm was proposed for multi-agent coordination. Swarm STL (SSTL) was proposed in~\cite{vande2023game} to express the swarm requirements and a game theoretic approach was proposed for distributed optimizations. 


Semantic robustness serves as an indicator to quantify the satisfaction degree of a specification with real values. The development of robustness metrics design and robustness maximization-based control methods constitute a pivotal research area, which has been extensively studied for bottom-up temporal logic. Various metrics, such as spatial robustness~\cite{smoothrobustness, leung2023backpropagation, dawson2022robust}, time robustness~\cite{9992914, rodionova2022combined}, and counting robustness~\cite{9929325} , have been proposed to evaluate satisfaction levels in bottom-up temporal logic tasks from spatial, temporal, and quantitative perspectives, respectively. However, literature addressing global temporal logic tasks remains relatively sparse, primarily focusing on spatial satisfaction \cite{leahy2021scalable, liu2023robust}. For instance, in tasks like multi-agent environmental monitoring, agents need to collaborate and complete tasks synchronously to enhance the effectiveness and accuracy of observations. In this case, evaluation metric is needed to quantify the quality of synchronization. Therefore, it is essential to explore a broader range of satisfaction metrics for global tasks to comprehensively meet diverse control requirements.

Motivated by these considerations, this paper focuses on the coordination control problem for multi-agent systems, aiming to satisfy global temporal logic requirements while considering agents with heterogeneous capabilities. We address complex global tasks that involve not only task completion timing, required agent capabilities, and repetition frequencies, but also coupling relationships between agents and  task interdependencies, such as sub-tasks that should be synchronously executed by agents. The tasks are formulated into global STL specifications. Considering the synchronous execution sub-tasks, we propose a metric, called synchronous robustness, to quantify the synchronization quality with real values. MILP methods are provided for encoding synchronous robustness, which employs a logarithmic number of binary variables and results in only linear complexity with respect to task length. Considering continuous workspace and discrete-time system dynamics, an optimization framework is presented to guarantee global task satisfaction with high synchronicity and minimal control efforts. Simulations demonstrate the efficiency of the motion planning strategy. 

This paper is structured as follows. In Section~\ref{section2}, we introduce STL and formulate the system model. Section~\ref{section3} introduces the syntax and semantics for global STL specifications. The control algorithm is detailed in Section~\ref{section5}. Simulation results are presented in Section~\ref{section6}, and Section~\ref{section7} concludes this paper.

\section{Preliminaries}\label{section2}
\subsection{Signal temporal logic}
The atomic predicate of the STL is defined as $\mu$, whose value is determined by a predicate function $ \alpha: \mathbb{R}^{n} \rightarrow \mathbb{R}$, i.e., $\mu$ is true if and only if $\alpha(x) \geq 0$ where $x\in\mathbb{R}^{n}$. The syntax of STL is defined recursively as
\begin{equation}\label{inner logic}
	\phi ::= \mu \mid \phi\wedge \varphi\mid \phi\vee \varphi\mid G_{[a, b]} \phi\mid F_{[a, b]} \phi \mid \phi U_{[a, b]} \varphi,
\end{equation}
where $\varphi$ and $\phi$ are all STL specifications;  $\wedge$, $\vee$ and $\neg$ are Boolean operators  conjunction, disjunction and negation, respectively; $[a,b], a, b \in \mathbb{Z}, a< b$ denotes a bounded time interval containing all time steps starting from $a$ to $b$; and $G_{[a, b]}$, $F_{[a, b]}$, $U_{[a, b]}$ are temporal operators \it globally\rm, \it finally \rm and \it until\rm, respectively. Specifically, $G_{[a, b]}\phi$ indicates that $\phi$ must hold continuously over the future time interval $[a,b]$, $F_{[a, b]} \phi$ means that $\phi$ must become true at least once within the interval $[a,b]$, and $\phi U_{[a, b]} \varphi$ requires that $\phi$ remains true until $\varphi$ becomes true within the interval $[a,b]$.  Let $\xi = x(0)x(1)\dots$ represent a discrete-time system run (signal), which is a sequence of system states where $x(k)$ denotes the state at time $k$. A signal $\xi$ satisfies $\phi$ at time $k$, denoted by $(\xi,k)\models\phi$, if the sequence $x(k)x(k+1)\dots$ satisfies $\phi$. The qualitative semantics of STL, which determines whether the STL task is satisfied or not, can be recursively defined as
\begin{equation*}\label{STL semantics}
	\begin{aligned}
		(\xi,k)\models\mu &\Leftrightarrow &&\!\!\! \alpha(x(k)) \geq 0 \\
		(\xi,k)\models\varphi_{1}\wedge \varphi_{2} &\Leftrightarrow && \!\!\! (\Xi,k)\models\varphi_{1}\wedge (\xi,k)\models\varphi_{2}\\
		(\xi,k)\models\varphi_{1}\vee \varphi_{2} &\Leftrightarrow && \!\!\! (\xi,k)\models\varphi_{1}\vee (\xi,k)\models\varphi_{2}\\
		(\xi,k)\models G_{[a,b]}\varphi &\Leftrightarrow && \!\!\! \forall k'\in [k+a,k+b],  (\xi,k')\models\varphi\\
		(\xi,k)\models F_{[a,b]}\varphi &\Leftrightarrow && \!\!\! \exists k'\in [k+a,k+b],  (\xi,k')\models\varphi\\
		(\xi,k)\models \varphi_{1}U_{[a,b]}\varphi_2&\Leftrightarrow&& \!\!\! \exists k'\in [k+a,k+b], \mathrm{s.\,t.\ } (\xi,k')\models\varphi_2 \wedge\forall k''\in [k,k'], (\xi,k'')\models\varphi_1.
	\end{aligned}
\end{equation*}

\subsection{System Model}
Consider a team of agents indexed by a finite set $\mathcal{P} = {1, 2, \dots, |\mathcal{P}|}$. Each agent $p \in \mathcal{P}$ is described by the tuple $\mathcal{A}_p = \langle\mathcal{X}_p, x_p(0), \mathcal{U}_p, Cap_p, f_p\rangle$. Here, $\mathcal{X}_p \subseteq \mathbb{R}^{n_x}$ and $\mathcal{U}_p \subseteq \mathbb{R}^{n_u}$ are the state and control spaces of the agent, respectively, with $x_p(0)$ representing its initial state. The set $Cap$ contains all possible capabilities, and $Cap_p \subseteq Cap$ specifies the capabilities of agent $p$. The dynamics of agent $p$ are governed by the function $f_p: \mathbb{R}^{n_x} \times \mathbb{R}^{n_u} \rightarrow \mathbb{R}^{n_x}$, following the relation:
\begin{equation*}
x_p(k+1) = A_px_p(k) +B_p u_p(k),
\end{equation*}
where $x_p(k) \in \mathcal{X}_p$ is the state and $u_p(k) \in \mathcal{U}_p$ is the control input at time step $k$. The matrices $A_p$ and $B_p$ are time-invariant. 

Given a fixed planning horizon $H \in \mathbb{N}_+$, which is determined by the task, the control input sequence for agent $p$ from time 0 to $H-1$ is denoted by $\mathbf{u}_p := [u_p^\top(0), u_p^\top(1), \dots, u_p^\top(H-1)]$. From the initial state $x_p(0)$ and using the control sequence $\mathbf{u}_p$, the resulting trajectory $\mathbf{s}_p := x_p(0) x_p(1) \dots x_p(H)$ can be obtained, which represents the states of agent $p$ over the time horizon $[0,H]$.

The collective trajectory of the entire agent team is represented by $\mathbf{S} := \{\mathbf{s}_p\}_{p \in \mathcal{P}}$. The subset of agents with a specific capability $C$ is denoted by $\mathcal{P}_C = \{p \mid C \in Cap_p\}$. An illustrative example of this agent model is provided in Example~\ref{example1}.

\section{Global tasks syntax and semantics}\label{section3}
A precision agriculture scenario is presented as follows to intuitively illustrate the tasks discussed in this paper.
\begin{example}\label{example1}
	\begin{figure}[tbp]
		\includegraphics[width=8.5cm]{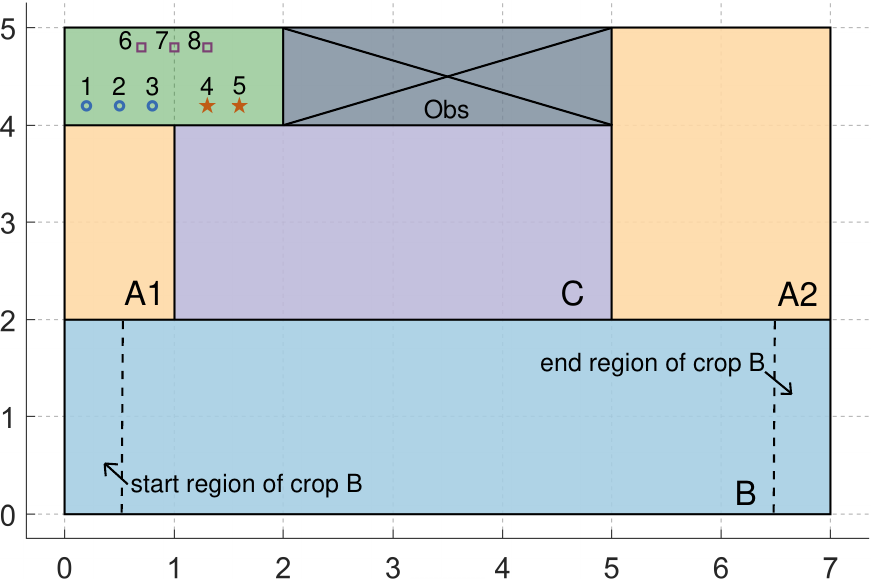}
		\centering
		\caption{A farmland health monitoring scenario. The initial position of each agent is shown in the upper left corner. Blue circles, orange pentagons, and purple squares represent agents equipped with UV, IR, and Vis sensors, respectively.}
		\label{map}
	\end{figure}
	Eight ground-based agents are tasked with monitoring farmland health, utilizing a variety of sensors: three ultraviolet (UV), two infrared (IR), and three visual (Vis) sensors. For the specified multi-agent system, we have $\mathcal{P} = \{1,2,\dots,8\}$, $Cap=\{UV, IR, Vis\}$, $\mathcal{P}_{UV}=\{1,2,3\}$, $\mathcal{P}_{IR}=\{4,5\}$, and $\mathcal{P}_{Vis}=\{6,7,8\}$. The dynamics of agent $p\in\mathcal{P}$ are given by 
	\begin{equation}
		\begin{aligned}
			\label{simudy}
			x_p(k+1) = \left[ \begin{matrix} I_2 & I_2\\0 & I_2\end{matrix}\right] x_p(k)+\left[\begin{matrix} 0.5 I_2\\I_2\end{matrix}\right]u_p(k),
		\end{aligned}
	\end{equation}
	where $x_p=\left[z_p^\top,v_p^\top\right]^\top$ with $z_p=[z_p^{x},z_p^{y}]^\top$ and  $v_p=[v_p^{x},v_p^{y}]^\top$ being the position and the velocity of agent $p$ in two directions, respectively, and control input $u_p=[u_p^{x}, u_p^{y}]^\top$.  The state space $\mathcal{X}_{p} = [0,7]\times[0,5]\times [-2,2]\times [-2,2]\subset \mathbb{R}^4$ illustrates the feasible domain for agent positions and velocities, and control space $\mathcal{U}_p=[-2, 2]\times [-2,2]\subset \mathbb{R}^2$. $I_2$ denotes the $2\times 2$ identity matrix. The map of the farm is shown in Fig.~\ref{map}, where each colored region corresponds to a different type of crop. The agents are initially situated within the initial area (the green region), and are required to monitor different crops while avoiding inaccessible regions (the gray region) and maintaining a safe distance from each other to prevent collisions. The monitoring requirements of each crop are described as follows. 
	
	1) Crop A: Within 7 time steps after the deployment, one UV sensor should stay in each subregion of A, including A1 and A2, for a duration of 2 time steps.
	
	2) Crop B: Between 2 to 7 time steps after the deployment, one UV and one IR sensor should form a team to traverse region B for collaborative data collection. Two such formations are required.
	
	3) Crop C: Within 3 to 7 time steps after the deployment, two Vis sensors should remain in region C for at least 1 time step simultaneously. In particular, the synchronization duration should be as long as possible to achieve complete observation of Crop~C.
\end{example}

In Example~\ref{example1}, the monitoring tasks for regions A, B, and C are complex global temporal logic tasks, which are imposed on agents with specific capabilities, and including requirements such as completion times and synchronous execution. Besides, considering the monitoring requirements for Crop C, how to achieve as long synchronization as possible should also be considered. In the following, we first formulate these group behaviors into global STL formulae, and then provide a formal robustness metric to evaluate the synchronicity.

\subsection{Syntax and qualitative semantics}
The global STL specification is based on a structure called \emph{task}. The \emph{task} incorporates an inner logic, which defines the STL task for individual or multiple coupled agents, and the \emph{task} specifies additional conditions for fulfilling the inner logic, such as necessary agent capabilities or the frequency of completion. Specifically, the syntax of the inner logic follows an STL specification \cite{bartocci2018specification}. Especially, the atomic predicate of inner logic is determined by a predicate function imposed on multiple agents, i.e., $\mu= (\alpha(x_{p^1},x_{p^2},\dots,x_{p^{N_\mu}}) \geq 0), \alpha:
\mathbb{R}^{n_x\times N_\mu} \rightarrow \mathbb{R}$, 
where $x_{p^{n}}, n\in[1, N_\mu], N_\mu\in\mathbb{N}$ denotes the state of the $n$th agent related to $\mu$. Note that $p^1,p^2, \dots p^{N_\mu}$ are agents from team~$\mathcal{P}$, but their specific identities are not defined.

The structure of \emph{task} is defined as $T = \langle \phi, c, Pattern \rangle$, which indicates that a \emph{task} $T$ is satisfied if and only if $\phi$ is satisfied for $c$ times following the requirements in $Pattern$. Specifically,

i) $\phi$ is an inner logic  defined by~\eqref{inner logic};

ii) $c\in\mathbb{N}^+$ is an integer that represents the number of times $\phi$ needs to be fulfilled;

iii) $Pattern$ indicates a set of constraints that indicates the criteria for agent selection. 

Let $N^\phi$ indicate the number of agents involved in inner logic $\phi$, and let $p_i^n\in\mathcal{P}, n \in [1,N^\phi], i\in[1,c]$ indicate the label of the $n$th agent in $\phi$ when $\phi$ is satisfied for the $i$th time. $Pattern$ then indicates constraints imposed on $p_i^n, n \in [1,N^\phi], i\in[1,c]$. For example, $p_i^n\in\mathcal{P}_{C}$ which restricts agent $p_i^n$ to be equipped with capability $C$, and $\bigcap_{i=1}^{c}\{p_i^1,\dots,p_i^{N_\phi}\} = \emptyset$ requires that no agent can be reused during the $c$-times-satisfaction of $\phi$. 

 With reference to the monitoring task for crop C in \emph{Example~1}, there may be synchronization requirements for the $c$-times-satisfaction of the inner logic (e.g., two Vis sensors should remain in region C for at least 1 time step simultaneously). To better indicate synchronous requirements, we construct the following \emph{synchronous} \emph{task} $T_s$ on the basis of general \emph{task}~$T$:
\begin{equation*}
	T_s:=F_{[a,b]} \langle G_{[0,d]}\varphi, c, Pattern\rangle.
\end{equation*}
In words, $T_s$ require $G_{[0,d]}\varphi$ to be satisfied for $c$ times during time interval $[a,b]$, and the $c$-times-satisfaction of $G_{[0,d]}\varphi$ needs to be achieved at the same time. Here, the inner logic in the \emph{synchronous} \emph{task} is restricted to the \it globally \rm manner, i.e., $G_{[0,d]}\varphi$ with $d\in\mathbb{N}$ and $\varphi$ being a specification given by \eqref{inner logic}. This is reasonable since synchronization requirements are generally imposed on persistent temporal logic tasks. Note that the valid time interval $[0,d]$ for formula $G_{[0,d]}\varphi$ always starts from time 0, as  $F_{[a,b]} \langle G_{[e,d+e]}\varphi, c, Pattern\rangle$ can always be rewritten into the form $F_{[a+e,b+e]} \langle G_{[0,d]}\varphi, c, Pattern\rangle$.

Based on the above statements, we formulate the global STL specification as follows:
\begin{equation}\label{global STL}
	\Phi ::=  T \mid T_s\mid \Phi_1\wedge\Phi_2,
\end{equation}
where $\Phi_1$ and $\Phi_2$ are all global STL formulae. Note that general STL formulae $\phi$ given by~\eqref{inner logic} can be regarded as special cases of \emph{task}s (where $c=1$ and the specific labels of the involved agents are clearly informed in the $Pattern$). Therefore, we unify these tasks within $\Phi$ and eliminate their distinct treatment below.

Recall the team trajectory $\mathbf{S}$. We say that  $\mathbf{S}$ satisfies task $\Phi$ at time $k$, denoted by $(\mathbf{S},k)\models\Phi$, if trajectory $\{x_{p}(k) x_{p}(k+1)\dots x_{p}(k+H)\}_{p\in\mathcal{P}}$ satisfies $\Phi$.  Accordingly, $\mathbf{S}$ satisfies $\Phi$, denoted by $\mathbf{S}\models\Phi$, if $(\mathbf{S},0)\models\Phi$. 

The qualitative semantics of task $\Phi$ provides a True or False value, meaning that the $\Phi$ is satisfied or not. Given agent tuples $\mathcal{A}_p$, $ p\in\mathcal{P}$ and a \emph{task} $T=\langle \phi, c, Pattern\rangle$, we can construct $N_{T}\in\mathbb{N}$ agent groups that meet the constraints in $Pattern$ using combinatorial techniques~\cite{riordan2014introduction}. Each group specifically comprises $c$ elements, and each element contains $N^\phi$ agents related with the inner logic~$\phi$. A detailed procedure for constructing the $N_{T}$ agent groups is presented in the following continuation of Example~\ref{example1}. Let $\mathbf{S}_{j}$ indicate the trajectory of agent group $j$, and let $\mathbf{S}_{j}^{i}$ indicate the trajectory of the $i$th element in group~$j$, $i\in[1,c]$, $j\in[1,N_{T}]$. The qualitative semantic of $T$ can then be given by 
\begin{equation}\label{qualitative semantic of task}
	(\mathbf{S},k)\models T \ \Leftrightarrow\ \bigvee_{j\in[1,N_{T}]}\{\bigwedge_{i\in[1,c]}[(\mathbf{S}_{j}^{i},k)\models \phi]\}.
\end{equation}
Here, the qualitative semantic of the involved inner logic, i.e., whether $(\mathbf{S}_{j}^{i},k)\models \phi$ holds,  is same as STL \cite{bartocci2018specification}. Let $N_{T_s}$ indicate the number of $Pattern$-satisfied agent groups for  \emph{synchronous task}  $T_s=F_{[a,b]} \langle G_{[0,d]}\varphi, c, Pattern\rangle$, which can be constructed similar to $T$ by solely considering $ \langle\varphi, c, Pattern\rangle$. The qualitative semantic of $T_s$ can be defined as 
\begin{equation*}
	(\mathbf{S},k)\models T_s  \ \Leftrightarrow \  \bigvee_{k'\in[a,b]}\{\bigvee_{j\in[1,N_{T_s}]}\{\bigwedge_{i\in[1,c]}[(\mathbf{S}_{j}^{i},k+k')\models G_{[0,d]}\varphi]\}\}. 
\end{equation*}
For the remaining conjunction operation $\wedge$ in~\eqref{global STL}, we can easily obtain the following:
\begin{equation}\label{qualitative semantic}
	(\mathbf{S},k)\models \Phi_1\wedge\Phi_2 \ \Leftrightarrow \ (\mathbf{S},k)\models\Phi_1\wedge(\mathbf{S},k)\models\Phi_2.
\end{equation}

\noindent\textbf{Example 1: }(continued) The monitoring tasks of crop regions can be formulated as global STL specifications as follows:
\begin{align}
	\Phi_{A1} = &\langle F_{[0,7]} G_{[0,2]} (z_{p^1}\in \mbox{A1}), 1, p_1^1\in\mathcal{P}_{UV} \rangle;\nonumber\\
	\Phi_{A2} = &\langle F_{[0,7]} G_{[0,2]} (z_{p^1}\in \mbox{A2}), 1, p_1^1\in\mathcal{P}_{UV} \rangle;\nonumber\\
	\Phi_B = &\langle F_{[0,2]}(z_{p^1},z_{p^2}\in \mbox{B}_{start})\wedge F_{[7,9]}(z_{p^1},z_{p^2}\in \mbox{B}_{end})\wedge G_{[2,7]}(z_{p^1},z_{p^2}\in \mbox{B})\wedge G_{[2,7]}(\|z_{p^1}-z_{p^2}\|_\infty\leq d_{form}),\nonumber\\ &2,\ \{p^1_1\in\mathcal{P}_{UV}, p^2_1\in\mathcal{P}_{IR}, p^1_2\in\mathcal{P}_{UV}, p^2_2\in\mathcal{P}_{IR}, \{p^1_1, p^2_1\}\cap \{p^1_2, p^2_2\}= \emptyset \} \rangle;\nonumber\\
	\Phi_C = &F_{[3,7]} \langle G_{[0,1]} (z_{p^1}\in \mbox{C}), 2, \{p^1_1\in\mathcal{P}_{Vis}, p^1_2\in\mathcal{P}_{Vis},  p^1_1 \neq p^1_2\} \rangle,\nonumber
\end{align}
where $d_{form}$ indicates a required formation distance. We use $\Phi_B$ as an example to demonstrate the design of agent groups. As depicted in Table~\ref{table1}, adhering to the constraints for agent selection in $Pattern$, we can establish six agent groups for fulfilling $\Phi_B$. This indicates that $\Phi_B$ can be accomplished by any of these six agent groups. Specifically, with $c=2$, each group comprises two elements. Given that the satisfaction of the inner logic in $\Phi_B$ necessitates $N^\phi=2$ agents, each element contains the labels of two agents. Regarding the collection of agent trajectories, we provide the following instances:  $\mathbf{S}_{3} = \{\mathbf{s}_1,\mathbf{s}_5, \mathbf{s}_2,\mathbf{s}_4\}$ and $\mathbf{S}_{3}^2 = \{ \mathbf{s}_2,\mathbf{s}_4\}$. Tasks imposed on specific agents, such as  obstacle and collision avoidance, can be modeled directly with STL specifications as follows: 
\begin{align}
	&\phi_{Obs} =  \bigwedge_{p\in\mathcal{P}} G_{[0,9]}(z_p\notin \mbox{Obs}),\nonumber\\
	&\phi_{Col} =  \bigwedge_{p\in\mathcal{P}} G_{[0,9]}(\bigwedge_{p'\in\mathcal{P}/p} \|z_{p}-z_{p'}\|_\infty\geq d_{col}),\nonumber
\end{align}
where $d_{col}$ indicates a required safe distance. These bottom-up tasks can be regarded as special cases of \emph{task}s and are unified within $\Phi$. The overall monitoring task can be formulated as $\Phi=\Phi_{A1}\wedge\Phi_{A2}\wedge\Phi_B\wedge\Phi_C\wedge\phi_{Obs}\wedge\phi_{Col}$.

\begin{table}
	\caption{Agent labels for task $\Phi_B$ under different agent groups}\label{table1}
	\begin{tabular*}{\tblwidth}{@{}CCCCCCC@{}}
		\toprule
		& group 1 ($j=1$) &group 2 ($j=2$)  &group 3 ($j=3$)  &group 4 ($j=4$) &group 5 ($j=5$)  &group 6 ($j=6$)  \\ 
		\midrule
		element 1 ($i=1$) & $p_1^1=1$, $p_1^2=4$ & $p_1^1=1$, $p_1^2=4$ & $p_1^1=1$, $p_1^2=5$ &$p_1^1=1$, $p_1^2=5$ & $p_1^1=2$, $p_1^2=4$ & $p_1^1=2$, $p_1^2=5$  \\
		element 2 ($i=2$)& $p_2^1=2$, $p_2^2=5$ & $p_2^1=3$, $p_2^2=5$ & $p_2^1=2$, $p_2^2=4$ & $p_2^1=3$, $p_2^2=4$ & $p_2^1=3$, $p_2^2=5$ & $p_2^1=3$, $p_2^2=4$ \\
		\bottomrule
	\end{tabular*}
\end{table}

\subsection{Synchronous robust semantics}\label{section4}
Robust semantics, known as robustness, measures how strongly a specification is satisfied with true values. In the following, we define a synchronous robustness for \emph{synchronous task} $T_s=F_{[a,b]} \langle G_{[0,d]}\varphi, c, Pattern\rangle$ to indicate the quality of synchronicity. Let $N_{T_s}$ indicate the number of $Pattern$-satisfied agent groups for $ T_s$, and let $[a_j,b_j],a_j,b_j\in\mathbb{N}, a_j\in[a,b],a_j < b_j$ represent the synchronization period for group~$j\in [1,N_{T_s}]$. Here the synchronization refers to that during time interval $[a_j,b_j]$, all trajectories $\mathbf{S}_j^i, i\in[1,c]$ in group $j$ satisfy $\varphi$. The robustness metrics of  $T_s$ are then designed as follows:
\begin{align}
	\rho_s(\mathbf{S},T_s,k)=&\max_{j\in[1,N_{T_s}]}(b_j-a_j)-d\label{Synchronous robustness}  \quad \mathrm{s.\,t.\ } a_j \in[a,b],\ (b_j-a_j)-d\geq0,\ d\in\mathbb{N}.
\end{align}
Here, constraints $a_j \in[a,b]$ and $\ (b_j-a_j)-d\geq0$ restrict that  the synchronous robustness is valid only for those trajectories that can satisfy $T_s$.

\begin{remark}
	The satisfaction of synchronous task can be evaluated based on various factors, such as the completion frequency of synchronous tasks and the duration of synchronization periods. In this paper, we focus on the scenarios where there are high demands for synchronization duration.
\end{remark}

Consider a global STL specification $\Phi$, which is assumed to be the conjunction of $M$ general \emph{task}s, denoted by $T^m= \langle \phi_m, c_m, Pattern_m\rangle$, $m=1,\dots,M$, and $L$ \emph{synchronous} \emph{task}s, denoted by $T_{s}^l= F_{[a_l,b_l]} \langle G_{[0,d_l]}\varphi_l, c_l, Pattern_l\rangle$, $l=1,\dots,L$, i.e., 
\begin{equation}\label{ECaTL task with subtasks}
	\Phi = \bigwedge_{m \in [1,M]}T^m \wedge\bigwedge_{l \in [1,L]}T_{s}^l.
\end{equation}
The synchronous robustness for $\Phi$ is given as
\begin{equation}\nonumber
	\rho_s(\mathbf{S},\Phi,k) = \min_{l\in [1,L]}\rho_s(\mathbf{S},T_s^l,k).
\end{equation}


\section{Control Synthesis}\label{section5}
In this section, we formulate and solve the robustness-maximized control synthesis problem for heterogeneous multi-agent systems under global STL tasks. 

\begin{problem}\label{problem1}
	Consider agents $\mathcal{A}_p, p\in\mathcal{P}$, an global STL task $\Phi$ as in \eqref{ECaTL task with subtasks}, and a planning horizon $H$. Find control input $\mathbf{u}_p$ for each agent $p\in\mathcal{P}$ such that the resulting team trajectory $\textbf{S}\models \Phi$ with~high synchronous robustness and minimal control efforts $\sum_{p\in\mathcal{P}}\|\mathbf{u}_p\|_2$ by solving the following optimization problem:
	\begin{align}
		\max_{\tilde{\mathbf{u}}_p, p\in\mathcal{P}}\ & \rho_s(\mathbf{S},\Phi,0) - \beta \sum_{p\in\mathcal{P}}\|\mathbf{\tilde{u}}_p\|_2\label{step1}\\
		\mathrm{s.\,t.\ } &  \mathcal{A}_p,\ \forall p\in\mathcal{P}, \tag{\ref{step1}a}\\
		& \mathbf{S}\models T^m,\ m=1,\dots,M, \tag{\ref{step1}b}\label{step1b} \\
		& 0\leq\rho_s(\mathbf{S},T_s^l,0)\leq \gamma^l,\ l=1,\dots,L, \tag{\ref{step1}c}\label{step1c}
	\end{align}where constraint~\eqref{step1b} enforces the satisfaction of \emph{task}s $T^m,m=1,\dots,M$, and the design of constraint~\eqref{step1c} and the objective function maximizes the synchronous robustness values for  \emph{synchronous} \emph{task}s $T^l,l=1,\dots,L$. Specifically, $\tilde{\mathbf{u}}_p$ indicates the control vector in Step~1, $\beta\in\mathbb{R}$ is a user-defined weighting parameter, and $\gamma^l\in\mathbb{N}$ indicates the upper bound for $\rho_s(\mathbf{S},T_s^l,0)$ of interest.
\end{problem}

To solve problem~\eqref{step1}, formula $\mathbf{S}\models T^m, m\in[1,M]$ and synchronous robustness $\rho_s(\mathbf{S},T_s^l,0)$ should be transformed into a manner understandable by optimization solvers, and better with low computational complexity. To achieve this, we introduce MILP encoding methods as follows, which employs a logarithmic number of binary variables and resulting in only linear complexity with respect to task length $H$.

\emph{1) MILP encoding for $(\mathbf{S},k)\models T$:}

The encoding aims to indicate whether $(\mathbf{S},k)\models T$ with indication variables and MILP constraints. The construction of these MILP constraints follows a recursive approach, with the specific encoding procedure for the involved inner logic $\phi $ and \emph{task} $T$ listed below.

The MILP encoding for inner logic $\phi$ is similar to STL. The basic idea is that for formula $\phi$ and a trajectory $(\mathbf{S}_{j}^{i},k)$, continuous indication variable $h[\mathbf{S}_{j}^{i},\phi,k]\in [0,1]$ is introduced such that $h[\mathbf{S}_{j}^{i},\phi,k]=1$ if the satisfaction of $\phi$ is enforced with $(\mathbf{S}_{j}^{i},k)$, and $h[\mathbf{S}_{j}^{i},\phi,k]\in [0,1)$ otherwise. The detailed encoding method can be found in \cite{9769752}.

The encoding for \emph{task} $T = \langle \phi, c, Pattern \rangle$ is based on its qualitative semantic~\eqref{qualitative semantic}. For team trajectory $(\mathbf{S}_{j},k)$, a continuous variable $h[\mathbf{S}_{j},T,k]\in [0,1]$ is introduced, and the following constraints are designed such that $h[\mathbf{S}_{j},T,k]=1$ enforces
$h[\mathbf{S}_{j}^{i},\phi,k]=1,\forall i\in[1,c]$, i.e., $(\mathbf{S}_{j},k)\models T$ holds:
\begin{align*}
	h[\mathbf{S}_{j}^{i},\phi,k]\geq h[\mathbf{S}_{j},T,k],\ \forall i\in[1, c].
\end{align*} Continuous indication variable $h[\mathbf{S},T,k]\in [0,1]$ is then employed and the following constraints are given such that if we set $h[\mathbf{S},T,k]=1$, then at least one $h[\mathbf{S}_{j},T,k],j\in[1,N_T]$ is enforced to be 1, i.e., at least one agent group $j\in[1,N_T]$ holds $(\mathbf{S}_{j},k)\models T$:
\begin{align*}
	\{1-h[\mathbf{S},T,k], h[\mathbf{S}_{1},T,k],\dots, h[\mathbf{S}_{N_{T}},T,k]\}\in SOS1,
\end{align*}
where $SOS1$ represents vectors that contain exactly one nonzero element, and that element is equal to~1. The detailed MILP encoding for $SOS1$ can be found in \cite{9769752}, which requires $\lceil \mathrm{log}_2(N_{T}+1) \rceil$ binary variables.

\emph{2) MILP encoding for $\rho_s(\mathbf{S},T_s,k)$:}

Recall the definition for synchronous robustness as in~\eqref{Synchronous robustness}, the key concept of the MILP encoding for $\rho_s(\mathbf{S},T_s,k)$, $T_s=F_{[a,b]} \langle G_{[0,d]}\varphi, c, Pattern\rangle$ is to capture the synchronous period $[a_j,b_j]$ of each group $j\in[1,N_{T_s}]$. The specific encoding procedure is as follows.

First, for each element $i$ in each group $j$, we introduce a set of continuous indication variables $h[\mathbf{S}_{j}^{i},\varphi,k']\in [0,1], k' = k+a,\dots, H$ and the MILP constraints as in \cite{9769752}, such that $h[\mathbf{S}_{j}^{i},\varphi,k']=1$ if the satisfaction of $\varphi$ is enforced with respect to $(\mathbf{S}_{j}^{i},k')$. 

We then encode the length of synchronous period $b_j-a_j$ into MILP constraints. With a slight abuse of notation, we define $\hat{T}_{s}: = \langle \varphi,c,Pattern \rangle$.  For each group $j$, continuous indication variables $h[\mathbf{S}_j, \hat{T}_{s},k']\in[0,1]$,  $k' = k+a,\dots, H$ are introduced with the following MILP constraints such that $h[\mathbf{S}_j, \hat{T}_{s},k']=1$ if the satisfaction of $\hat{T}_{s}$ is enforced with $(\mathbf{S}_j,k')$:
\begin{align*}
	h[\mathbf{S}_{j}^{i},\varphi,k']\geq h[\mathbf{S}_{j},\hat{T}_{s},k'],\ \forall i\in[1, c].
\end{align*}
We further limit $h[\mathbf{S}_j, \hat{T}_{s},k'],k'\in [k+a,H]$ as binary variables by
\begin{align*}
	h[\mathbf{S}_j, \hat{T}_{s},k'] = q^{k'}_j, \ q^{k'}_j\in\{0,1\},
\end{align*}
which will be utilized in the subsequent counting operations.
Counting variables $c[\mathbf{S}_j,\hat{T}_{s},k']\in\mathbb{R}$, $k'\in [k+a,H]$ are then constructed for each group $j$ to indicate the maximum number of sequential $h[\mathbf{S}_j, \hat{T}_{s},k'']=1$ when $k''\geq k'$. This is achieved by the recursively-defined counting constraints as follows:
\begin{align}\label{product}
	c[\mathbf{S}_j,\hat{T}_{s},k'] = (c[\mathbf{S}_j,\hat{T}_{s},k'+1]+1)\cdot h[\mathbf{S}_j, \hat{T}_{s},k'],\ c[\mathbf{S}_j,\hat{T}_{s},H+1]=0.
\end{align}

The biggest element in the first $b-a+1$ counting variables indicates the synchronization duration of group $j$, that is
\begin{equation}\label{choose time robustness}
	b_j-a_j = \max_{k'\in[k+a, k+b]}c[\mathbf{S}_j,\hat{T}_{s},k'].
\end{equation}
The synchronous robustness $\rho_s(\mathbf{S},T_s,k)$ can then be represented as
\begin{align} \rho_s(\mathbf{S},T_s,k)=\max_{j\in[1,N_{T_s}]}\max_{k'\in[k+a, k+b]}c[\mathbf{S}_j,\hat{T}_{s},k']-d.\label{max}
\end{align}
If $l$ synchronous \it task\rm s are included in $\Phi$ as in \eqref{ECaTL task with subtasks}, we have 
\begin{equation}\label{total constraint}
	\rho_s(\mathbf{S},\Phi,k) = \min_{l\in[1,L]}\rho_s(\mathbf{S},T_s^l,k).
\end{equation}

\begin{remark}Constraint \eqref{product} involves a product of integer and binary variables and constraints \eqref{choose time robustness}, \eqref{max} and \eqref{total constraint} involve max and min operations. Methods in \cite{BEMPORAD2001382} show that such product and min, max operations can be expressed into MILP constraints.
\end{remark}

In the MILP encoding rules described above, disjunctive formulae are encoded only using a logarithmic number of binary variables with respect to task length, while predicates and conjunctive formulae can be encoded without binary variables. We can then transform problem~\eqref{step1} into an MILP, which can be solved efficiently with the art-to-state solvers like Gurobi\footnote{http://www.gurobi.com}.

\section{Simulation}\label{section6}
\begin{figure}[htb]
	\centering
	\subfigure[Case I (the original task scenario).]{\includegraphics[width=7.2cm]{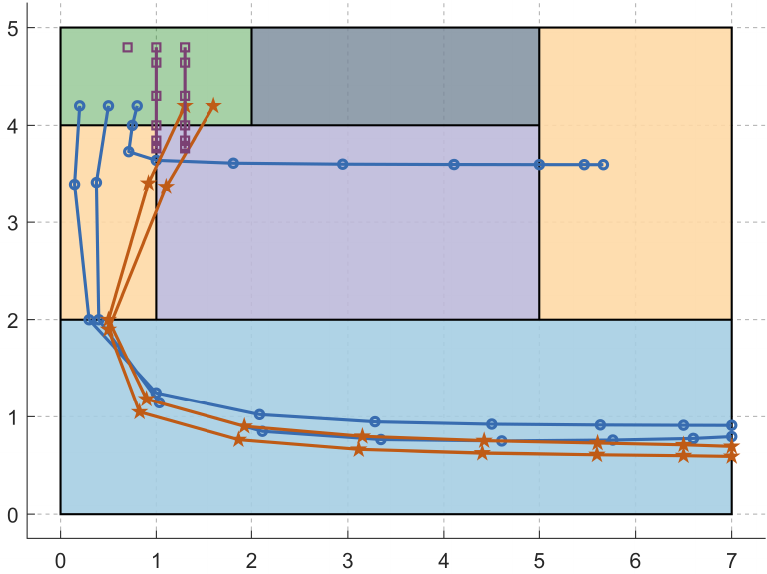}}\quad
	\subfigure[Case II.]{\includegraphics[width=7.2cm]{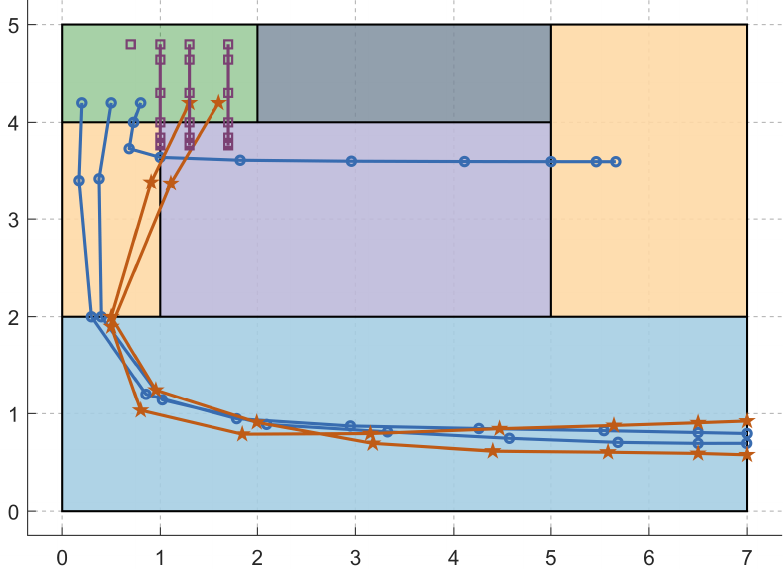}}
	\subfigure[Case III.]{\includegraphics[width=7.2cm]{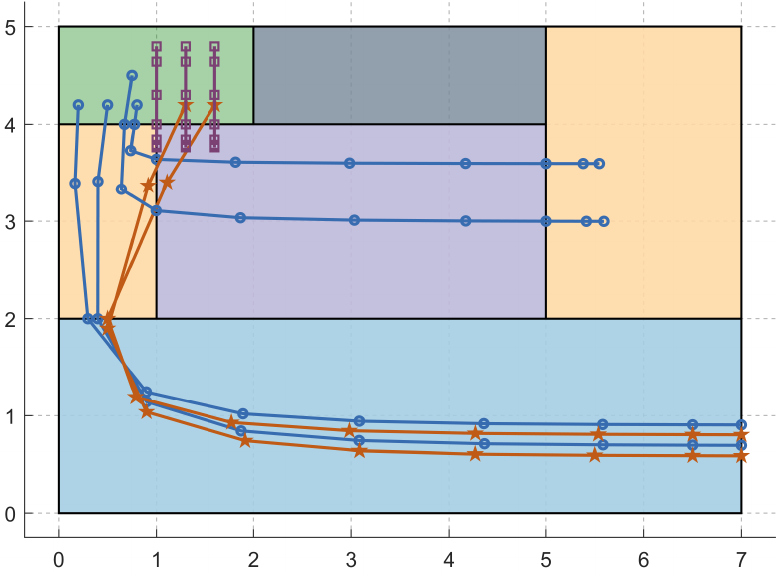}}
	\caption{Agent trajectories under the proposed motion planning algorithm }\label{agent trajectory1}
\end{figure}

In this section, we evaluate our control strategy using the precision agriculture scenario described in Example 1. All simulations are implemented in Matlab on a laptop with an AMD R9 5900HS 3.30 GHz processor and 16 GB RAM. Let $H = 9$, $\beta = 0.05$, $\gamma^1 = 8$, $d_{form} = 0.25$,  and $d_{col}=0.05$. We apply Gurobi to solve the MILP optimization problem.  From time 0, the resulting trajectory for each agent under the proposed control strategies is shown in Fig.~\ref{agent trajectory1}(a). One can see that all agents arrive at the monitoring crops within the specified time period with collision and obstacle avoidance.  As the synchronous duration is maximized, the synchronous time of $\Phi_C$ is determined as 7 from time step~3. 

\begin{table}
	\caption{Summary statistics  for Simulations.}\label{table3}
	\begin{tabular*}{\tblwidth}{@{}CCCC@{}}
		\toprule
		&\textbf{Task scenario}&\textbf{Computation time (mean / max)}  \\ 
		\midrule
		&Case I &  256.25 / 460.98   \\
		&Case II & 662.87 / 2085.66  \\
		&Case III & 897.67 / 2307.85   \\
		\bottomrule
	\end{tabular*}
\end{table}

The scalability of the proposed two-step algorithm is verified with the following task scenarios:

Case I: Original task scenario as in Example 1;

Case II: Increase the number of Vis agents to $|\mathcal{P}_{Vis}| = 4$ and raise the $c$ in $\Phi_C$ to 3;

Case III: Increase the number of UV agents to $|\mathcal{P}_{UV}| = 4$  and raise the $c$ in $\Phi_{A1}$ and $\Phi_{A2}$ to 2.

We conducted 30 simulations for each case, where the initial positions of agents were randomly selected within the initial region while meeting collision avoidance requirements. Specific agent trajectories under Case~I, Case~II, and Case~III are provided in Figure~\ref{agent trajectory1}. The computation time, including mean and maximum values, from these simulations is presented in Table~\ref{table3}. One can observe that, despite increasing both the number of agents and task completion times $c$, the proposed strategy achieves task satisfaction with synchronous robustness, indicating a certain level of scalability. It is worth noting that, as the computation time increases with the problem complexity, developing more efficient problem-solving strategies could be a future research direction.

\section{Conclusion}\label{section7}
This paper addresses the multi-agent control problem under global temporal logic tasks, focusing on agents with heterogeneous capabilities. We formalize tasks into global STL formulae, specifically accounting for aspects such as agent capabilities, repetition frequencies, agent coupling, and task interdependencies like synchronous execution. To evaluate synchronization quality, the concept of synchronous robustness is introduced, and a MILP encoding method is then proposed to facilitate motion planning with high synchronicity and minimal control effort. Future work includes investigating more expressive syntax and robustness, as well as developing efficient online control strategies.

\section{Acknowledgments}
This research was supported by the National Natural Science Foundation of China under Grant Nos.~62173224 and 92367203.

\section{Conflict of interest}

The authors declare no potential conflict of interests.












\printcredits

\bibliographystyle{cas-model2-names}

\bibliography{wileyNJD-AMA}

\begin{thebibliography}{26}
\expandafter\ifx\csname natexlab\endcsname\relax\def\natexlab#1{#1}\fi
\providecommand{\url}[1]{\texttt{#1}}
\providecommand{\href}[2]{#2}
\providecommand{\path}[1]{#1}
\providecommand{\DOIprefix}{doi:}
\providecommand{\ArXivprefix}{arXiv:}
\providecommand{\URLprefix}{URL: }
\providecommand{\Pubmedprefix}{pmid:}
\providecommand{\doi}[1]{\href{http://dx.doi.org/#1}{\path{#1}}}
\providecommand{\Pubmed}[1]{\href{pmid:#1}{\path{#1}}}
\providecommand{\bibinfo}[2]{#2}
\ifx\xfnm\relax \def\xfnm[#1]{\unskip,\space#1}\fi
\bibitem[{Bai et~al.(2022)Bai, Zheng, Xu, Liu and Zhang}]{bai2022hierarchical}
\bibinfo{author}{Bai, R.}, \bibinfo{author}{Zheng, R.}, \bibinfo{author}{Xu,
  Y.}, \bibinfo{author}{Liu, M.}, \bibinfo{author}{Zhang, S.},
  \bibinfo{year}{2022}.
\newblock \bibinfo{title}{Hierarchical multi-robot strategies synthesis and
  optimization under individual and collaborative temporal logic
  specifications}.
\newblock \bibinfo{journal}{Robotics and Autonomous Systems}
  \bibinfo{volume}{153}, \bibinfo{pages}{104085}.
\bibitem[{Bartocci et~al.(2018)Bartocci, Deshmukh, Donz{\'e}, Fainekos, Maler,
  Ni{\v{c}}kovi{\'c} and Sankaranarayanan}]{bartocci2018specification}
\bibinfo{author}{Bartocci, E.}, \bibinfo{author}{Deshmukh, J.},
  \bibinfo{author}{Donz{\'e}, A.}, \bibinfo{author}{Fainekos, G.},
  \bibinfo{author}{Maler, O.}, \bibinfo{author}{Ni{\v{c}}kovi{\'c}, D.},
  \bibinfo{author}{Sankaranarayanan, S.}, \bibinfo{year}{2018}.
\newblock \bibinfo{title}{Specification-based monitoring of cyber-physical
  systems: a survey on theory, tools and applications},
  \bibinfo{publisher}{Springer}. pp. \bibinfo{pages}{135--175}.
\bibitem[{Bemporad et~al.(2001)Bemporad, Torrisi and Morari}]{BEMPORAD2001382}
\bibinfo{author}{Bemporad, A.}, \bibinfo{author}{Torrisi, F.},
  \bibinfo{author}{Morari, M.}, \bibinfo{year}{2001}.
\newblock \bibinfo{title}{Discrete-time hybrid modeling and verification of the
  batch evaporator process benchmark}.
\newblock \bibinfo{journal}{European Journal of Control} \bibinfo{volume}{7},
  \bibinfo{pages}{382--399}.
\bibitem[{Buyukkocak and Aksaray(2022)}]{9992914}
\bibinfo{author}{Buyukkocak, A.T.}, \bibinfo{author}{Aksaray, D.},
  \bibinfo{year}{2022}.
\newblock \bibinfo{title}{Temporal relaxation of signal temporal logic
  specifications for resilient control synthesis}, pp.
  \bibinfo{pages}{2890--2896}.
\bibitem[{Buyukkocak et~al.(2021)Buyukkocak, Aksaray and
  Yaz{\i}c{\i}o{\u{g}}lu}]{buyukkocak2021planning}
\bibinfo{author}{Buyukkocak, A.T.}, \bibinfo{author}{Aksaray, D.},
  \bibinfo{author}{Yaz{\i}c{\i}o{\u{g}}lu, Y.}, \bibinfo{year}{2021}.
\newblock \bibinfo{title}{Planning of heterogeneous multi-agent systems under
  signal temporal logic specifications with integral predicates}.
\newblock \bibinfo{journal}{IEEE Robotics and Automation Letters}
  \bibinfo{volume}{6}, \bibinfo{pages}{1375--1382}.
\bibitem[{Dawson and Fan(2022)}]{dawson2022robust}
\bibinfo{author}{Dawson, C.}, \bibinfo{author}{Fan, C.}, \bibinfo{year}{2022}.
\newblock \bibinfo{title}{Robust counterexample-guided optimization for
  planning from differentiable temporal logic}, \bibinfo{organization}{IEEE}.
  pp. \bibinfo{pages}{7205--7212}.
\bibitem[{Gilpin et~al.(2021)Gilpin, Kurtz and Lin}]{smoothrobustness}
\bibinfo{author}{Gilpin, Y.}, \bibinfo{author}{Kurtz, V.},
  \bibinfo{author}{Lin, H.}, \bibinfo{year}{2021}.
\newblock \bibinfo{title}{A smooth robustness measure of signal temporal logic
  for symbolic control}.
\newblock \bibinfo{journal}{IEEE Control Systems Letters} \bibinfo{volume}{5},
  \bibinfo{pages}{241--246}.
\bibitem[{Gundana and Kress-Gazit(2021)}]{gundana2021event}
\bibinfo{author}{Gundana, D.}, \bibinfo{author}{Kress-Gazit, H.},
  \bibinfo{year}{2021}.
\newblock \bibinfo{title}{Event-based signal temporal logic synthesis for
  single and multi-robot tasks}.
\newblock \bibinfo{journal}{IEEE Robotics and Automation Letters}
  \bibinfo{volume}{6}, \bibinfo{pages}{3687--3694}.
\bibitem[{Kurtz and Lin(2022)}]{9769752}
\bibinfo{author}{Kurtz, V.}, \bibinfo{author}{Lin, H.}, \bibinfo{year}{2022}.
\newblock \bibinfo{title}{Mixed-integer programming for signal temporal logic
  with fewer binary variables}.
\newblock \bibinfo{journal}{IEEE Control Systems Letters} \bibinfo{volume}{6},
  \bibinfo{pages}{2635--2640}.
\bibitem[{Leahy et~al.(2021)Leahy, Serlin, Vasile, Schoer, Jones, Tron and
  Belta}]{leahy2021scalable}
\bibinfo{author}{Leahy, K.}, \bibinfo{author}{Serlin, Z.},
  \bibinfo{author}{Vasile, C.I.}, \bibinfo{author}{Schoer, A.},
  \bibinfo{author}{Jones, A.M.}, \bibinfo{author}{Tron, R.},
  \bibinfo{author}{Belta, C.}, \bibinfo{year}{2021}.
\newblock \bibinfo{title}{Scalable and robust algorithms for task-based
  coordination from high-level specifications (scratches)}.
\newblock \bibinfo{journal}{IEEE Transactions on Robotics}
  \bibinfo{volume}{38}, \bibinfo{pages}{2516--2535}.
\bibitem[{Leung et~al.(2023)Leung, Ar{\'e}chiga and
  Pavone}]{leung2023backpropagation}
\bibinfo{author}{Leung, K.}, \bibinfo{author}{Ar{\'e}chiga, N.},
  \bibinfo{author}{Pavone, M.}, \bibinfo{year}{2023}.
\newblock \bibinfo{title}{Backpropagation through signal temporal logic
  specifications: Infusing logical structure into gradient-based methods}.
\newblock \bibinfo{journal}{The International Journal of Robotics Research}
  \bibinfo{volume}{42}, \bibinfo{pages}{356--370}.
\bibitem[{Li et~al.(2022)Li, Cai, Xiao and Kan}]{9687668}
\bibinfo{author}{Li, Z.}, \bibinfo{author}{Cai, M.}, \bibinfo{author}{Xiao,
  S.}, \bibinfo{author}{Kan, Z.}, \bibinfo{year}{2022}.
\newblock \bibinfo{title}{Online motion planning with soft metric interval
  temporal logic in unknown dynamic environment}.
\newblock \bibinfo{journal}{IEEE Control Systems Letters} \bibinfo{volume}{6},
  \bibinfo{pages}{2293--2298}.
\bibitem[{Lindemann et~al.(2019)Lindemann, Nowak, Sch{\"o}nb{\"a}chler, Guo,
  Tumova and Dimarogonas}]{lindemann2019coupled}
\bibinfo{author}{Lindemann, L.}, \bibinfo{author}{Nowak, J.},
  \bibinfo{author}{Sch{\"o}nb{\"a}chler, L.}, \bibinfo{author}{Guo, M.},
  \bibinfo{author}{Tumova, J.}, \bibinfo{author}{Dimarogonas, D.V.},
  \bibinfo{year}{2019}.
\newblock \bibinfo{title}{Coupled multi-robot systems under linear temporal
  logic and signal temporal logic tasks}.
\newblock \bibinfo{journal}{IEEE Transactions on Control Systems Technology}
  \bibinfo{volume}{29}, \bibinfo{pages}{858--865}.
\bibitem[{Liu et~al.(2023)Liu, Leahy, Serlin and Belta}]{liu2023robust}
\bibinfo{author}{Liu, W.}, \bibinfo{author}{Leahy, K.},
  \bibinfo{author}{Serlin, Z.}, \bibinfo{author}{Belta, C.},
  \bibinfo{year}{2023}.
\newblock \bibinfo{title}{Robust multi-agent coordination from catl+
  specifications}, \bibinfo{organization}{IEEE}. pp.
  \bibinfo{pages}{3529--3534}.
\bibitem[{Luo and Zavlanos(2022)}]{luo2022temporal}
\bibinfo{author}{Luo, X.}, \bibinfo{author}{Zavlanos, M.M.},
  \bibinfo{year}{2022}.
\newblock \bibinfo{title}{Temporal logic task allocation in heterogeneous
  multirobot systems}.
\newblock \bibinfo{journal}{IEEE Transactions on Robotics}
  \bibinfo{volume}{38}, \bibinfo{pages}{3602--3621}.
\bibitem[{Maler and Nickovic(2004)}]{STL_intro}
\bibinfo{author}{Maler, O.}, \bibinfo{author}{Nickovic, D.},
  \bibinfo{year}{2004}.
\newblock \bibinfo{title}{Monitoring temporal properties of continuous
  signals}, pp. \bibinfo{pages}{152--166}.
\bibitem[{Qin et~al.(2016)Qin, Ma, Shi and Wang}]{qin2016recent}
\bibinfo{author}{Qin, J.}, \bibinfo{author}{Ma, Q.}, \bibinfo{author}{Shi, Y.},
  \bibinfo{author}{Wang, L.}, \bibinfo{year}{2016}.
\newblock \bibinfo{title}{Recent advances in consensus of multi-agent systems:
  A brief survey}.
\newblock \bibinfo{journal}{IEEE Transactions on Industrial Electronics}
  \bibinfo{volume}{64}, \bibinfo{pages}{4972--4983}.
\bibitem[{Qiu et~al.(2024)Qiu, Meng, Xu and Yang}]{10458342}
\bibinfo{author}{Qiu, X.}, \bibinfo{author}{Meng, W.}, \bibinfo{author}{Xu,
  J.}, \bibinfo{author}{Yang, Q.}, \bibinfo{year}{2024}.
\newblock \bibinfo{title}{Reduced-order observer-based resilient control for
  mass with time-varying delay against dos attacks}.
\newblock \bibinfo{journal}{IEEE Transactions on Industrial Cyber-Physical
  Systems} \bibinfo{volume}{2}, \bibinfo{pages}{51--59}.
\bibitem[{Ren et~al.(2024)Ren, Long, Ma and Li}]{10574325}
\bibinfo{author}{Ren, H.}, \bibinfo{author}{Long, Y.}, \bibinfo{author}{Ma,
  H.}, \bibinfo{author}{Li, H.}, \bibinfo{year}{2024}.
\newblock \bibinfo{title}{Distributed group coordination of random
  communication constrained cyber-physical systems using cloud edge computing}.
\newblock \bibinfo{journal}{IEEE Transactions on Industrial Cyber-Physical
  Systems} \bibinfo{volume}{2}, \bibinfo{pages}{196--205}.
\bibitem[{Riordan(2014)}]{riordan2014introduction}
\bibinfo{author}{Riordan, J.}, \bibinfo{year}{2014}.
\newblock \bibinfo{title}{An introduction to combinatorial analysis} .
\bibitem[{Rodionova et~al.(2022)Rodionova, Lindemann, Morari and
  Pappas}]{rodionova2022combined}
\bibinfo{author}{Rodionova, A.}, \bibinfo{author}{Lindemann, L.},
  \bibinfo{author}{Morari, M.}, \bibinfo{author}{Pappas, G.J.},
  \bibinfo{year}{2022}.
\newblock \bibinfo{title}{Combined left and right temporal robustness for
  control under stl specifications}.
\newblock \bibinfo{journal}{IEEE Control Systems Letters} \bibinfo{volume}{7},
  \bibinfo{pages}{619--624}.
\bibitem[{Sahin et~al.(2019)Sahin, Nilsson and Ozay}]{sahin2019multirobot}
\bibinfo{author}{Sahin, Y.E.}, \bibinfo{author}{Nilsson, P.},
  \bibinfo{author}{Ozay, N.}, \bibinfo{year}{2019}.
\newblock \bibinfo{title}{Multirobot coordination with counting temporal
  logics}.
\newblock \bibinfo{journal}{IEEE Transactions on Robotics}
  \bibinfo{volume}{36}, \bibinfo{pages}{1189--1206}.
\bibitem[{Vande~Kamp et~al.(2023)Vande~Kamp, Koochakzadeh, Yazicioglu and
  Aksaray}]{vande2023game}
\bibinfo{author}{Vande~Kamp, L.}, \bibinfo{author}{Koochakzadeh, A.},
  \bibinfo{author}{Yazicioglu, Y.}, \bibinfo{author}{Aksaray, D.},
  \bibinfo{year}{2023}.
\newblock \bibinfo{title}{A game theoretic approach to distributed planning of
  multi-agent systems under temporal logic specifications}, p.
  \bibinfo{pages}{1657}.
\bibitem[{Yang et~al.(2023)Yang, Zou, Li and Yang}]{yang2023distributed}
\bibinfo{author}{Yang, T.}, \bibinfo{author}{Zou, Y.}, \bibinfo{author}{Li,
  S.}, \bibinfo{author}{Yang, Y.}, \bibinfo{year}{2023}.
\newblock \bibinfo{title}{Distributed model predictive control for
  probabilistic signal temporal logic specifications}.
\newblock \bibinfo{journal}{IEEE Transactions on Automation Science and
  Engineering, doi: 10.1109/TASE.2023.3323472} .
\bibitem[{Zhang and Gong(2023)}]{10207769}
\bibinfo{author}{Zhang, W.}, \bibinfo{author}{Gong, B.}, \bibinfo{year}{2023}.
\newblock \bibinfo{title}{Observer-based consensus of nonlinear multi-agent
  systems with input delay via delay-dependent event-triggered control}.
\newblock \bibinfo{journal}{IEEE Transactions on Industrial Cyber-Physical
  Systems} \bibinfo{volume}{1}, \bibinfo{pages}{157--166}.
\bibitem[{Zhou et~al.(2023)Zhou, Yang, Zou, Li and Fang}]{9929325}
\bibinfo{author}{Zhou, X.}, \bibinfo{author}{Yang, T.}, \bibinfo{author}{Zou,
  Y.}, \bibinfo{author}{Li, S.}, \bibinfo{author}{Fang, H.},
  \bibinfo{year}{2023}.
\newblock \bibinfo{title}{Multiple subformulae cooperative control for
  multiagent systems under conflicting signal temporal logic tasks}.
\newblock \bibinfo{journal}{IEEE Transactions on Industrial Electronics}
  \bibinfo{volume}{70}, \bibinfo{pages}{9357--9367}.

\end{thebibliography}



\end{document}